\documentclass[conference]{IEEEtran}

\usepackage{blindtext}
\usepackage{float}
\usepackage{graphicx}
\graphicspath{{/home/panos/Documents/l}}

\usepackage[belowskip=-5pt,aboveskip=5pt]{caption}
 
\begin{document}

\title{A Data Imputation Model based on an Ensemble Scheme}

\author{\IEEEauthorblockN{Panagiotis Fountas}
\IEEEauthorblockA{\textit{Department of Informatics and Telecommunications} \\
\textit{University of Thessaly}\\
Lamia, Greece  \\
email: pfountas@uth.gr}
\and
\IEEEauthorblockN{Kostas Kolomvatsos}
\IEEEauthorblockA{\textit{Department of Informatics and Telecommunications} \\
\textit{University of Thessaly}\\
Lamia, Greece \\
email: kostasks@uth.gr}
}


\maketitle

\IEEEpubidadjcol

\begin{abstract}
Edge Computing (EC) offers an infrastructure that acts as the mediator between the Cloud and the Internet of Things (IoT). The goal is to reduce the latency that we enjoy when relying on Cloud. IoT devices interact with their environment to collect data relaying them towards the Cloud through the EC. Various services can be provided at the EC for the immediate management of the collected data. One significant task is the management of missing values. In this paper, we propose an ensemble based approach for data imputation that takes into consideration the spatio-temporal aspect of the collected data and the reporting devices. We propose to rely on the group of IoT devices that resemble to the device reporting missing data and enhance its data imputation process. We continuously reason on the correlation of the reported streams and efficiently combine the available data. Our aim is to `aggregate' the local view on the appropriate replacement with the `opinion' of the group. We adopt widely known similarity techniques and a statistical modelling  methodology to deliver the final outcome. We provide the description of our model and evaluate it through a high number of simulations adopting various experimental scenarios.
\end{abstract}

\section{Introduction}
The Internet of Things (IoT) consists of a huge infrastructure where numerous devices can interact with users and their environment to collect and process data.
IoT devices create streams of data towards the Cloud infrastructure where advanced processing takes place.
Between the Cloud and the IoT, one can meet the Edge Computing (EC)
where various nodes with different computational capabilities are present.
These nodes can perform the first stage of processing for the data that IoT devices collect.
For instance, simple processing activities like data filtering, 
novelty detection, etc or simple analytics tasks can be realized at the EC.
It becomes obvious that 
the discussed processing can support the necessary services to eliminate the latency in responses to effectively deal with real time applications.
Computing and processing of data are now close to the IoT devices
giving the opportunity to perform 
analytics over the distributed streams. 
We have to notice that the discussed data streams are geo-located towards the EC nodes
where distributed datasets are also formulated to become the subject of
the aforementioned processing activities. 

When focusing on data processing, one major research problem is the management of missing values.
Any fault in the collected data may jeopardize the quality of outcomes of any processing activity \cite{he}.
Researchers have already focused on this problem and proposed 
a set of models to support the efficient provision of replacements \cite{guan}.
Some of the proposed techniques involve data exclusion, missing indicator analysis, mean substitution, single imputation, multiple imputation techniques, replacement at random, etc.
All of them conclude to specific formulations to deliver the replacements for every missing value.
They try to respond to the critical research question of how we can manage/replace a missing value when observed in a data stream.
The majority of the proposed schemes deals with the statistical information that a stream `conveys' adopting it to estimate the most probable value to replace the missing one. 
Obviously, we have to detect the hidden distribution relying behind the collected data, then, to 
easily replace the missing value.

The current paper consists of the continuation of our previous work in the field and advances the state of the art by proposing an ensemble based approach to perform the envisioned data imputation scheme.
Initially, we propose to use a statistical learning method to estimate 
missing values based on the `experience' of the device reporting them.
We rely on a linear regression model \cite{freedman} due to its simplicity and the ability to be adopted 
to support real time applications.
Our aim is to estimate missing values based on past observations. 
Additionally, we rely on a `group' oriented approach, i.e., we get similar
IoT devices to contribute on the estimation of missing values. Through the term `similar', we denote devices that exhibit the same spatio-temporal characteristics with the device reporting 
every missing value. 
We consider devices that are in a close distance and report similar data for the phenomenon under consideration.
Hence, we are able to support a group- and data-aware model that results the final replacements. 
We also perform an `aggregation' of the `local' view,
i.e., the view of the device reporting the missing value on the replacements with the view of the group and propose a dynamic adaptation mechanism for the weight adopted to deliver the final outcome.
Our model can be easily incorporated in a monitoring mechanism placed at the EC being directly connected with IoT devices
to perform the envisioned data imputation on the fly. 
The aforementioned similarity between devices and their reports is realized by the known cosine similarity \cite{han} and the Mahalanobis distance \cite{han} to expose the correlation between data streams.
We have to notice that this is a continuous process, thus, we rely on a sliding window approach to focus only on the the most recent reports. 
The Mahalanobis distance is adopted to 
`calibrate' the results retrieved by the Cosine similarity applied
over the most recent reports of IoT devices.
Eventually, we are able to detect the similarity between
devices' reports relying over their current and past behaviour.
Both, the Mahalanobis distance and the Cosine Similarity are widely used 
in many types of problems. They are characterized by simplicity and 
the ability of providing fast results which is critical for (near) real time 
applications. 
The differences of the current approach with our previous efforts \cite{fountas} are as follows:
\begin{itemize}
 	\item we extend our previous model and adopt a statistical modelling process, i.e., a regression analysis to estimate missing values based on the past observations of devices;
 	\item we combine the `local' view of each device with the view of the group of devices based on a dynamic adaptation scheme for weights adopted to conclude the aggregation of the proposed replacements; 
 	\item we adopt the geometrical mean for aggregating the `opinion' of devices participating in the group of similar peers.
\end{itemize} 
 
The remaining paper is organized as follows. Section \ref{section2} reports on the related work while Section \ref{section3} presents the problem under consideration and gives insights into our model. Section \ref{section4} discusses the proposed solution and provides the relevant formulations. Section \ref{section5} presents our experimental evaluation efforts and gives numerical results for outlining the pros and cons of our model. Finally, in Section \ref{section6}, we conclude our paper and describe our future
research plans in the domain.

\section{Related work}
\label{section2}
Offering novel applications in IoT is gaining significant attention 
in recent years. The reason is that these applications are provided
close to end users, thus, we can enjoy real time responses.
Any application is performed over data, i.e., a data processing activity that 
is adopted to create knowledge in various forms \cite{escamilla}.
Due to the presence of numerous devices,
the volumes of the collected data becomes huge \cite{cai}. 
However, IoT devices usually exhibit limited computational capabilities that are not appropriate to perform complex processing activities.
On the other side, Cloud offers a vast processing infrastructure but when we rely on it,
we enjoy increased latency in the provision of responses.
The research community proposed the adoption of EC and Fog Computing to limit the latency by 
incorporating processing nodes with enhanced computational capabilities (compared to IoT devices) 
close to IoT and users.
A comparison between EC, 
Fog computing (FC), cloudlets and mobile EC is presented in 
\cite{dolui}. 
The authors provide a comparative analysis of the three implementations  
together with the necessary parameters that affect nodes 
communication (e.g., physical proximity, 
access mediums, context awareness, power consumption, computation time).

Data storage technologies become the means for setting up the basis to perform the desired 
processing tasks \cite{ruiz}.
Any storage mechanism should be optimized to decide the necessary models that maximize the 
access rate to data and the performance in general.
The stored data can be structured or unstructured \cite{jiang}.
This means that, when processing should take place, we have to 
combine multiple data sources (e.g., databases).
The authors of \cite{habak} propose a system to facilitate  
and support a set of services at the edge of the network.
They envision a set of clusters and adopt a controller to add devices to these clusters, thus, the system can perform a resource allocation and assign 
the desired processing tasks.
The discussed storage models should exhibit the 
necessary security levels to avoid any malfunctions or unauthorized access.
The blockchain technology can assist towards this direction \cite{shafagh}.
In \cite{fu}, the authors present a scheme for security management in an IoT data storage system
incorporating a data pre-processing task realized at the EC.
Another distributed data storage mechanism is provided by \cite{xing}.
The authors propose a multiple factor replacement algorithm 
to manage the limited storage resources and data loss. 

A critical part in the pre-processing of the collected data 
is the necessary processing for data imputation.
The aim is to efficiently handle possible missing values.
The specific research subject is widely studied by the research community 
and a large of technologies have been proposed.
These techniques are adopted in various applications domains exhibiting its 
significance for building novel applications.
In some of the first attempts, data imputation has been adopted to manage sensory data 
\cite{jiangnan}. Researchers propose the use of statistical metrics for concluding the 
replacements of missing values.
The simplest approach is to replace a missing value with the
mean of the incoming samples. 
However, the adoption of the mean cannot take into consideration 
the variance of data or their correlation
\cite{little} being also affected by extreme values.
To take into consideration data correlations, more advanced statistical learning 
schemes can be also utilized. This way, we can 
incorporate the  
statistical dependencies of data into our decision making mechanism
\cite{ku}, \cite{zhao}. 
Example models can be found in \cite{chang}, i.e., Auto-Regressive Integrated Moving Average 
and the feed forward prediction based method. 
The aforementioned statistical models are applied over 
historical values, thus, we have to take into consideration 
the processing time for concluding the final outcome.
Apart of the time requirements, our decision making should 
also take into account, the prediction error that may be present.
Moreover, to focus on recent data, we can rely on a sliding window approach aiming to reduce the processing time and be aligned with `fresh' information.
Probabilistic approaches focus on the extraction of the distribution of data adopted 
to `generate' the final replacement 
\cite{zhao}.
Other efforts 
deal with the joint distribution on the entire data set
assuming a parametric
density functions (e.g., a multivariate normal) on the data given with estimated
parameters 
\cite{honaker}. 
The technique of least squares
provides individual univariate regressions to impute features with missing values
on all of the other dimensions based on the weighted average of 
the individual predictions 
\cite{bo}, \cite{Raghunathan}.
Extensions of the least squares method consist of the
Predictive-Mean Matching method (PMM) where replacements are random
samples drawn from a set of observed values close to regression predictions
\cite{Buuren}
and 
Support Vector Regression (SVR) 
\cite{wang}.
Other imputation models involve 
random forests 
\cite{Stekhoven},
K-Nearest Neighbors (K-NN) 
\cite{Troyanskaya},
sequential K-NN
\cite{Kim}, 
singular value decomposition and 
linear combination of a set of eigenvectors
\cite{Troyanskaya}, \cite{Mazumder} and
Bayesian Principal Component
Analysis (BPCA)
\cite{Mohamed}, \cite{Oba}.
Probabilistic Principal Component Analysis (PPCA) and 
Mixed Probabilistic Principal Component Analysis (MPPCA) 
can be also adopted to impute data
\cite{zhao}.
All the aforementioned techniques try to deal with data that are not linearly correlated
providing a more `generic' model.
Formal optimization can be also adopted to impute missing data with mixed continuous and categorical
variables \cite{bertsimas}. 
The optimization model incorporates various predictive models and can 
be adapted for multiple imputations. 

It becomes obvious that the definition of replacements 
involves an uncertainty about the final adopted value.
This uncertainty can be managed by technologies like Fuzzy Logic (FL).
The aim is to avoid adopting crisp thresholds when deciding the 
final replacement.
Machine learning (ML) and Computational Intelligence (CI) can also 
assist in the analysis of data and the replacement of missing values
opening the room for defining intelligent schemes.
Some example efforts are as follows. A hybrid method that adopts the Fuzzy C-means (FCM) algorithm
combined with a Particle Swarm Optimization (PSO) model and a Support Vector Machine (SVM)
is presented in \cite{Shang}. 
Other models involve Multi-layer
Perceptrons (MLPs) \cite{Reznik}, 
Self-Organizing Maps (SOMs) \cite{Catterall},
and Adaptive Resonance Theory (ART) \cite{Carpenter}. 
The use of Neural Networks (NN) can lead to 
capture many kinds of relationships among data and they allow quick
and easy modelling of the environment \cite{Li}. 

\section{Preliminaries \& Problem Description}
\label{section3}
Our scenario involves a set of IoT devices that interact with end users and their environment collecting data or performing simple processing activities. The description of the scenario `borrows' the notation of \cite{fountas} as both papers deal with the same problem. Any processing at the devices targets to produce knowledge locally, close to end users.   
These devices can be directly connected through the network with edge nodes to transfer the collected data in an upwards mode. The final 
goal is to transfer data to the Cloud infrastructure. 
Edge nodes act as `sinks' that store the information provided by the IoT devices (see Figure 1 \cite{fountas}). 
We consider that devices report multivariate vectors in the following form
$\mathbf(x) = \left\langle x_{1}, x_{2}, \ldots, x_{M} \right\rangle$ where $M$ represents the number of dimensions.
Edge nodes can convey the necessary components to fuse the incoming vectors and store them in the appropriate format to become the subject of further processing.
Each vector is annotated by the index of the $j$th device reporting it to the corresponding edge node and the reporting time instance $t$,
i.e., $\mathbf(x)^{j}[t] = \left\langle x_{1}^{j}[t], x_{2}^{j}[t], \ldots, x_{M}^{j}[t] \right\rangle$. 
Without loss of generality, we consider that $N$ devices are connected
with an edge node.
After the reception of data, edge nodes should perform the desired pre-processing to prepare the data to become the subject of various tasks. In this paper, our focus is on the processing of missing values and the application of a data imputation model. 
We propose the use of multiple technologies to conclude the replacement for each missing value. 
We have to notice that missing values can refer in entire vectors or
in the inputs for specific dimensions.

We also focus on the most recent measurements reported by IoT devices, i.e., we consider the $W$ latest reports (see Table I). 
Actually, in Table I, we describe the two-dimensional structure for 
the $j$th individual IoT device.
It is a strategic decision to incorporate in our model only fresh information. Any data received out of the window $W$ is considered as  obsolete. This is a typical scheme of a sliding window approach.
$W$ deals with the interval where
data can be adopted to assist in the delivery of missing values 
replacements.

\begin{figure}[!h]
\label{fig1}
	\centering
	\includegraphics[scale=0.15]{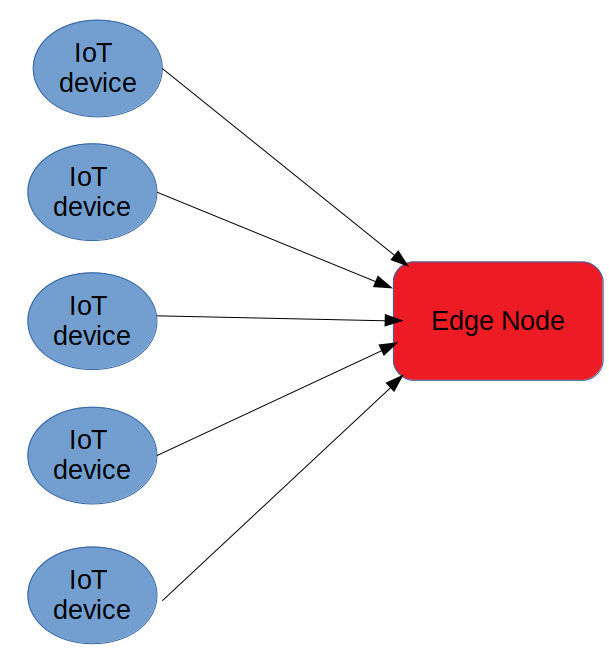}
	\caption{The architecture considered in our scenario.}
\end{figure}

\begin{table}[!h]
\label{table1}
\caption{An example of the collected IoT devices reports.}
\centering
\begin{tabular}{|c ||c | c | c | c||}
\hline
&1st dimension & 2nd dimension & ... & $M$th dimension \\ [1ex] 
\hline
t=1&$x_1^{j}[1]$ & $x_2^{j}[1]$ & ... & $x_M^{j}[1]$  \\[1ex] 
\hline
t=2&$x_1^{j}[2]$ & $x_2^{j}[2]$ & ... & $x_M^{j}[2]$  \\[1ex] 
\hline
... & ... & ... & ...& ... \\
\hline
t=W& $x_1^{j}[W]$ & $x_2^{j}[W]$ & ... &$x_M^{j}[W]$  \\ [1ex] 
\hline
\end{tabular}
\end{table}

The proposed model is `fired' when a missing value is detected.
Then, we provide the means for calculating the replacement based on the local `experience', i.e., the previous reports of the same device where a missing value is detected and the experience of the group, i.e., 
the replacement `proposed' by the group of devices exhibiting similar data in the spatio-temporal axis.
Edge nodes rely on an ensemble scheme that identifies the 
correlation between data reported by devices combining 
current and historical information defined in $W$.
Our model consists of the following parts:
(i) a module responsible to apply a linear regression model and estimate the replacement based on the local knowledge/data (past data reported by the device where a missing value is detected);
(ii) a module that realizes the correlations of the collected vectors from all the available devices;
(iii) an ensemble scheme to detect the final correlation between IoT devices as `exposed' by their reports;
(iv) a module that produces the final replacement taking into consideration the local view (the outcome of the regression process) as well as the view of peer device (devices with increased similarity with the device reporting the missing value) adopting a weighted scheme;
(v) a module that dynamically adapts the weights for the local and the group view based on the deviation of reports of the device where the missing value is present. The deviation is calculated in the window $W$.
For the replacement `proposed' by the peer devices, we intent to rely only on data reported by correlated devices assuming that 
their spatio-temporal contextual information will positively affect the
final calculations.
For instance, if IoT devices monitor the same phenomenon in the same 
area, they should report similar values.
Hence, any missing information in individual reports can be easily 
managed adopting the `view' of the remaining devices in the group.

\section{Data Imputation based on Data Streams Correlation}
\label{section4}

\subsection{The envisioned setup}
Our Prediction Based Model (PBM) adopts an ensemble scheme to extract the correlation between the IoT devices upon the following popular metrics: 
(i) the Cosine Similarity (CS) and (ii) the Mahalanobis Distance (MD). 
The use of the aforementioned metrics does not differ in the PBM compared to our previous 
effort, i.e., the Distance Based Model (DBM) \cite{fountas}. 
The CS is adopted to calculate the similarity between the last reported data 
vectors between two different devices. Additionally, the MD targets to calculate the multi-dimensional distance between the reports of different IoT devices. In other words, we can say that the CS is applied at the `vector' level while the MD is applied at the `device' level.
We have to notice that the latter metric considers the entire `history' of reports as depicted by the latest $W$ vectors.
The outcomes of the CS and MD metrics are smoothly combined to deliver the final correlation between IoT devices reports, thus, to proceed with the calculation of the replacement values. 
For achieving this, we consider the top-$k$ similarity outcomes based on the devices correlation.

\subsection{Similarity between data streams}
The CS consists of a simple, however, powerful metric for the delivery of similarity between two two non-empty vectors. The metric is defined as the cosine of the angle between them. The CS is particularly used in positive space where the cosine of the angle is bounded in [0,1] and this value is inversely proportional to the angle between the vectors. So, the higher the angle between them, the lower the similarity becomes. 
The following function holds true:
\begin{equation} 
\label{eq:1}
\begin{array}{l}
CS(\mathbf(x)^{i}[t], \mathbf(x)^{j}[t]) = \frac{\mathbf(x)^{i}[t] \cdot \mathbf(x)^{j}[t]}{\|\mathbf(x)^{i}[t]\| \cdot \|\mathbf(x)^{j}[t]\|} = \\ 
\frac{\sum_{l=1}^{M}\mathbf(x)^{i}_{l}[t] \mathbf(x)^{j}_{l}[t]}{\sqrt{\sum_{l=1}^{M}(\mathbf(x)^{i}_{l}[t])^{2}} \cdot \sqrt{\sum_{l=1}^{M}(\mathbf(x)^{i}_{l}[t])^{2}}}
\end{array}
\end{equation}
In Eq(\ref{eq:1}), $i$ \& $j$ are the indexes of vectors/IoT devices fed into the 
CS function.
CS is applied over pairs of IoT devices for a specific time instance $t$, i.e., the latest/current report.

The CS gives us a view on the devices reporting similar values for the phenomenon under consideration. When a missing values is detected, we perform set of calculations, i.e., we feed the CS function with the 
latest reports (i.e., vectors) of the available devices. 
For simplifying calculations and avoid any undesired errors,
we do not take into consideration the dimension(s) where missing value is present. 
The CS is applied for the remaining dimensions of the discussed vectors. 

The MD measures the correlation between multivariate vectors as the distance between an observation point (which can be a set of observations or a single value) and a distribution.
Another case of the MD application is to detect the correlation between two multivariate vectors delivered by the same dataset.
Let $\overrightarrow{x}$ and $\overrightarrow{y}$ be two multivariate vectors reported by two IoT devices.
The following equation holds true:
\begin{equation}
MD(\overrightarrow{x}-\overrightarrow{y}) =	\sqrt{(\overrightarrow{x}-\overrightarrow{y})^{T}S^{-1}(\overrightarrow{x}-\overrightarrow{y})}
\end{equation}

In our case, the MD is applied on pairs of devices over the $W$ latest vectors that each IoT device reports to the edge node. Using this approach, we pay attention, not only on the latest reports but also on historical values of the devices. The results that will arise from the application of MD function are combined with the CS results to calculate the final correlation result.

\subsection{Local estimation of imputed values}
For the presentation of the approach, let us focus on a specific dimension considering, without loss of generality, that the incoming values are depicted by the variables $x[1], x[2], \ldots, x[W]$.
Let us consider that at a time instance $W+1$, we observe a missing value, thus, we have to estimate $x[W+1]$.
We consider $x[W+1]$ as the dependent variables and try to detect the linear relationship between $x[1], x[2], \ldots, x[W]$ and $x[W+1]$.
The `typical' approach is to adopt 
the following equation:
$x[W+1] = f\left( X, B \right) + \epsilon = b_{0} + b_{1}x[1]+ b_{2}x[2]+ \ldots, b_{W}x[W] + \epsilon$ where $b_{0}$  is the intercept, $b_{i}$ are the weights for each independent values
and $\epsilon$ is the error.
Our aim is to find the appropriate weights $\left\lbrace b_{i}, i=0,1,2,\ldots,W \right\rbrace$ that minimize the sum of squared errors, i.e., $\sum \left( x[W+1] - f\left( X, B \right)  \right)^{2}$. The least squares method is widely adopted because the estimated function $f\left( X, \hat{B} \right)$ approximates the conditional expectation $\mathbf{E}(x[W+1]|X)$. At this point, we omit the performed calculations that are widely studied in the research community. The interested reader can refer in \cite{freedman} for further details. In any case, adopting the discussed calculations we can easily retrieve $x[W+1]$ combining it with the proposed replacement by the group of similar devices.

\subsection{Our ensemble approach} 
In order to combine the correlation metrics as exposed by the CS and the MD, we adopt a simple data-aware correlation model that uses the MD to 'calibrate' the result of the CS. Hence, we create a correlation scheme based on the correlation exposed by the latest reports being affected by their historical `course'. More specifically, every result of the CS is weighted (i.e., multiplied) by $w$ defined as $w = \frac{1}{MD}$. Our the goal, by adopting this weight, is to assess a reward on IoT devices with high historical correlation. Hence, when the MD outcome increases, we observe a decrement of the final $w$ realization, i.e., we eliminate the weight, thus, the final correlations between two different devices. 
It is a strategic decision adopted in our model to pay less attention 
into IoT devices with low historical correlation (i.e., a high MD) regardless of the correlation detected by the CS. On the other hand, a low MD value indicates a strong historical correlation, thus, 
we `reward' the similarity of the latest vectors exposed by the CS. In some manner, the 
MD metric confirms or rejects the correlation depicted by the latest reports.

The final correlation outcome $F_{C}$ is derived by the following equation:
\begin{equation}
F_{C} = w \cdot CS(\mathbf(x)^{i}[t], \mathbf(x)^{j}[t]), \forall i,j, i \neq j
\end{equation}

Afterwards, we rely on the top-$k$ $F_{C}$ outcomes. 
The IoT devices that correspond to the top-$k$ values formulate the group of the most similar reports.
The discussed group becomes the basis for the calculation of the final replacement for each observed missing values.
The replacements are calculated taking into consideration the group `view' and the `local' view. For aggregating the view of each device participating in the group,
we adopt the Weighted Geometric Mean (WGM) \cite{Jung}. 
The WGM is calculated as follow:
\begin{equation}
\label{WGM}
    WGM= \left(\prod _{i=1}^{k}x_{i}^{MD_{i}}\right)^{\frac{\mathrm{1}}{\sum_{i=1}^{k} MD_{i}}}
\end{equation}
In Eq(\ref{WGM}), $x_{i}$ represents the report of every top-$k$ correlated device for the specific dimension where a missing value is detected and $MD_{i}$ is the MD between the IoT device in which we detect the missing value with the $i$th top-$k$ correlated device.

We perform a set of calculations for delivering $F_{C}$, WGM and the estimated value delivered by the linear regression model, i.e., $x[W+1]$. For aggregating the `group' with the `local' view, we adopt a dynamic weighted scheme upon the WGM and the $x[W+1]$. Actually, we propose a heuristic for calculating 
the weight of $x[W+1]$ as follows:

\begin{equation}
\label{eq:sigmoid}
w_{local} = \frac{\mathrm{1} }{\mathrm{1} + e^ {\alpha \sigma - \beta}}
\end{equation}
where $\sigma$ depicts the deviation of the $W$ latest reports of the device where a missing values is detected.
For the realization of $w_{local}$, we rely on the aforementioned sigmoid function, i.e.,
we eliminate the weight of the `local' view when $\sigma$ is over a predefined threshold. 
The rationale is that when $\sigma$ is high, there are disturbances in the distribution of data,
thus, we cannot support an efficient decision making based only on the `local' view.
Then, $w_{local}$ is retrived to be close to zero. 
Our PBM algorithm adopts $w_{local}$ and calculates the final replacement value as follows:

\begin{equation}
PD =  w_{local} \cdot x[W+1] + (1 - w_{local}) \cdot WGM
\end{equation}

It becomes obvious that the final $PD$ is the result of an ensemble scheme that combines multiple metrics and builds upon the view of the team of similar devices (as exposed by their data) and the 
view (exposed by historical reports) of the device where a missing value is present. 
The interesting is that we rely on a dynamic scheme for selecting the top-$k$ similar devices and take into consideration the distribution of data 
present in the dimension where the missing value is observed. 

\section{Experimentation Setup \& Performance Assessment}
\label{section5}

\subsection{Experimental Setup and Performance Metrics} 
We report on the performance of the proposed model related to its ability to correctly replace the detected missing values. Our experimental evaluation relies on two real traces, i.e.,
(i) the GNFUV Unmanned Surface Vehicles Sensor Data Set \cite{harth} and 
(ii) the Intel Berkeley Research Lab dataset
\footnote{Intel Lab Data, http://db.csail.mit.edu/labdata/labdata.html}.
The former dataset (i.e., the GNFUV) consists of values of mobile sensor readings  (humidity, temperature) from four Unmanned Surface Vehicles (USVs) moving in the sea according to a GPS predefined trajectory. 
The Intel dataset contains millions of measurements (temperature, humidity, light) retrieved 
by 54 sensors deployed in a lab. 
From this dataset, we get 15,000 measurements such that 15 sensors produced 1,000 measurements. The aforementioned traces are adopted to simulate the streams reported by a set of IoT devices.

The evaluation of the PBM is performed in two axes, i.e., its ability to eliminate  the prediction error and the capability of 
reducing the time required to conclude 
a replacement. 
The prediction error is estimated by the difference of the final outcome and the  actual value provided any of the above datasets. For the second axis, we rely on a wide set of experiments and get the mean time to depict the performance of the model. 
Our experimentation involves the `creation' of missing values in the available datasets by randomly annotating $V$\% reports (i.e., values in our datasets), from the total, as missing inputs. 
The following table presents the realization 
of each parameter as adopted into our experimentation evaluation. 

\begin{table}[!h]
\label{table2}
\caption{Realization of our Parameters}
\centering
\begin{tabular}{|c ||c | c | c | c||}
\hline
Short Description & Values \\ [1ex] 
\hline
Percentage of missing values & $V=\left\lbrace1,5,10\right\rbrace$  \\[1ex] 
\hline
Size of sliding window & $W = 10$   \\[1ex] 
\hline
Number of correlated devices adopted in our model & $k=4$   \\ [1ex] 
\hline
Total number of IoT devices & $N \in \left\lbrace 5,7,15 \right \rbrace$ \\ [1ex] 
\hline
Number of dimensions & $M=\left\lbrace 4,9 \right\rbrace$ \\[1ex]
\hline
Parameters adopted by our smoothing function & $\alpha=$20$, \beta=2$ \\[1ex]
\hline
\end{tabular}
\end{table}

The performance of our model is delivered through widely adopted 
metrics like the Mean Absolute Error (MAE) and the Root Mean Square Error (RMSE). The following equations hold true:
\begin{equation} \label{eq:5}
MAE=\frac{\sum_{l=1}^{n}|PD_{l} - a_{l}|}{n}
\end{equation}

\begin{equation} \label{eq:6}
RMSE=\sqrt{\frac{\sum_{l=1}^{n}(PD_{l} - a_{l})^{2}}{n}}
\end{equation}
In Eq(\ref{eq:5}) and Eq(\ref{eq:6}), $n$ is the number of missing values, $PD_{l}$ is the value predicted by our model and $a_{l}$ is the actual value as registered in the adopted datasets.
We also compare our scheme with two other models, i.e., a scheme proposed in one of our previous research efforts, i.e., the DBM \cite{fountas} and a baseline model, i.e., 
a model that replaces missing values with the mean 
of the incoming reports for the specific dimension where a missing value is observed. The mean is calculated upon the devices exhibiting 
a high correlation with the device reporting a missing input.
We name this model as the \textit{Averaging Model} (AM). Through the above evaluation process we try to find out if the proposed model is capable to support real time applications while providing efficient results in estimation of missing values.

\subsection{Performance Assessment}
\begin{figure}[!h]
	\centering
	\includegraphics[scale=0.10]{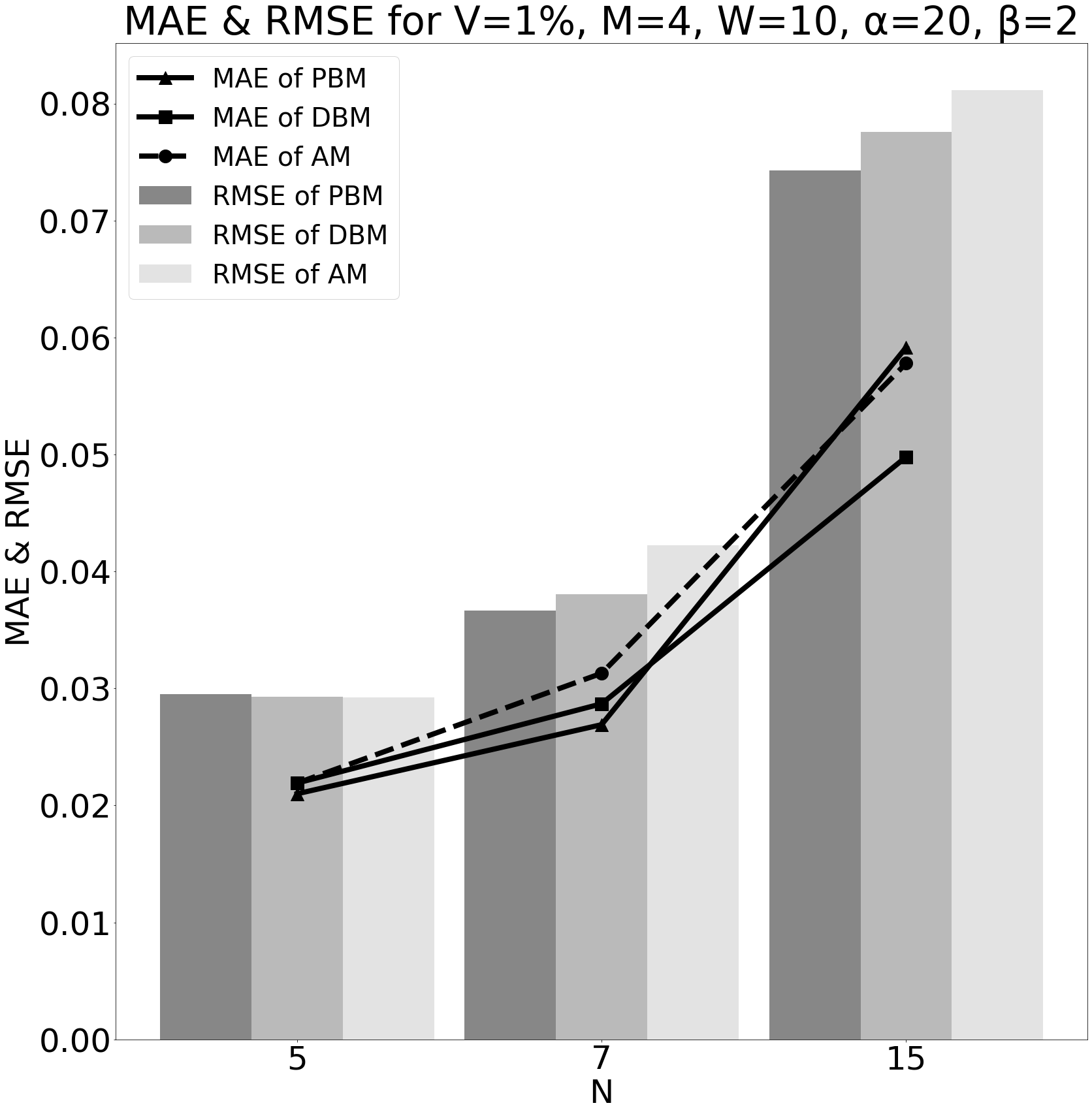}
	\caption{MAE and RMSE for $V = 1$\% and $M = 4$. }
	\label{fig2}
\end{figure}

\begin{figure}[!h]
	\centering
	\includegraphics[scale=0.10]{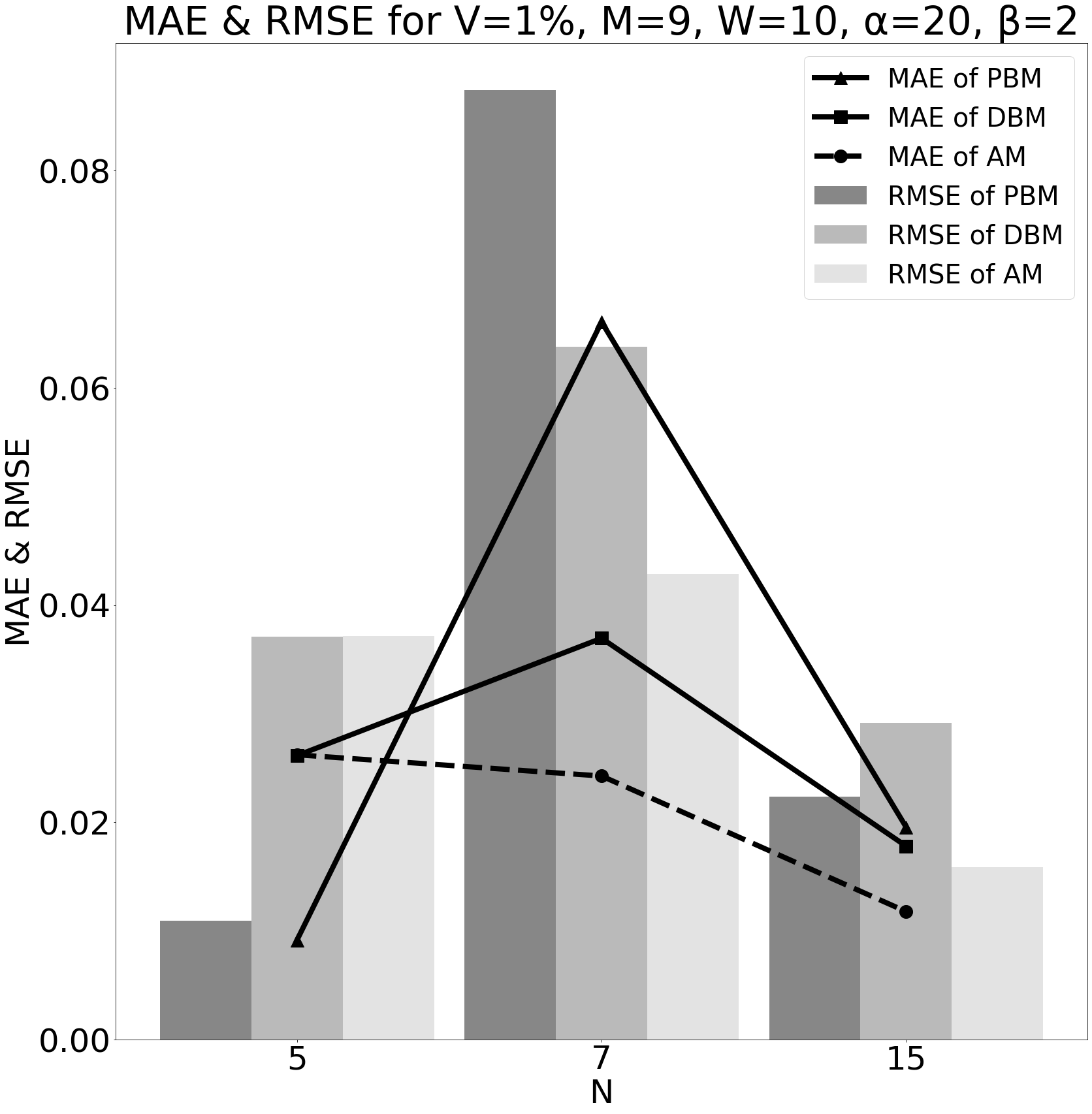}
	\caption{MAE and RMSE for $V = 1$\% and $M = 9$.}
	\label{fig3}
\end{figure}

In Figures \ref{fig2} \& \ref{fig3}, we present our results for $M=4$ and $M=9$, respectively when $V=1$\%.
As we can observe, the performance of the PBM is affected by the number of devices for both experimental scenarios resulting an increment in the error realizations.
The DBM and the AM are also negatively affected by the increment in the number of dimensions and the number of devices.
The comparative assessment reveals that the PBM performs better (except one case, i.e,. $M=9$, $N=7$ ) than the remaining models. 

\begin{figure}[!h]
\centering
	\includegraphics[scale=0.10]{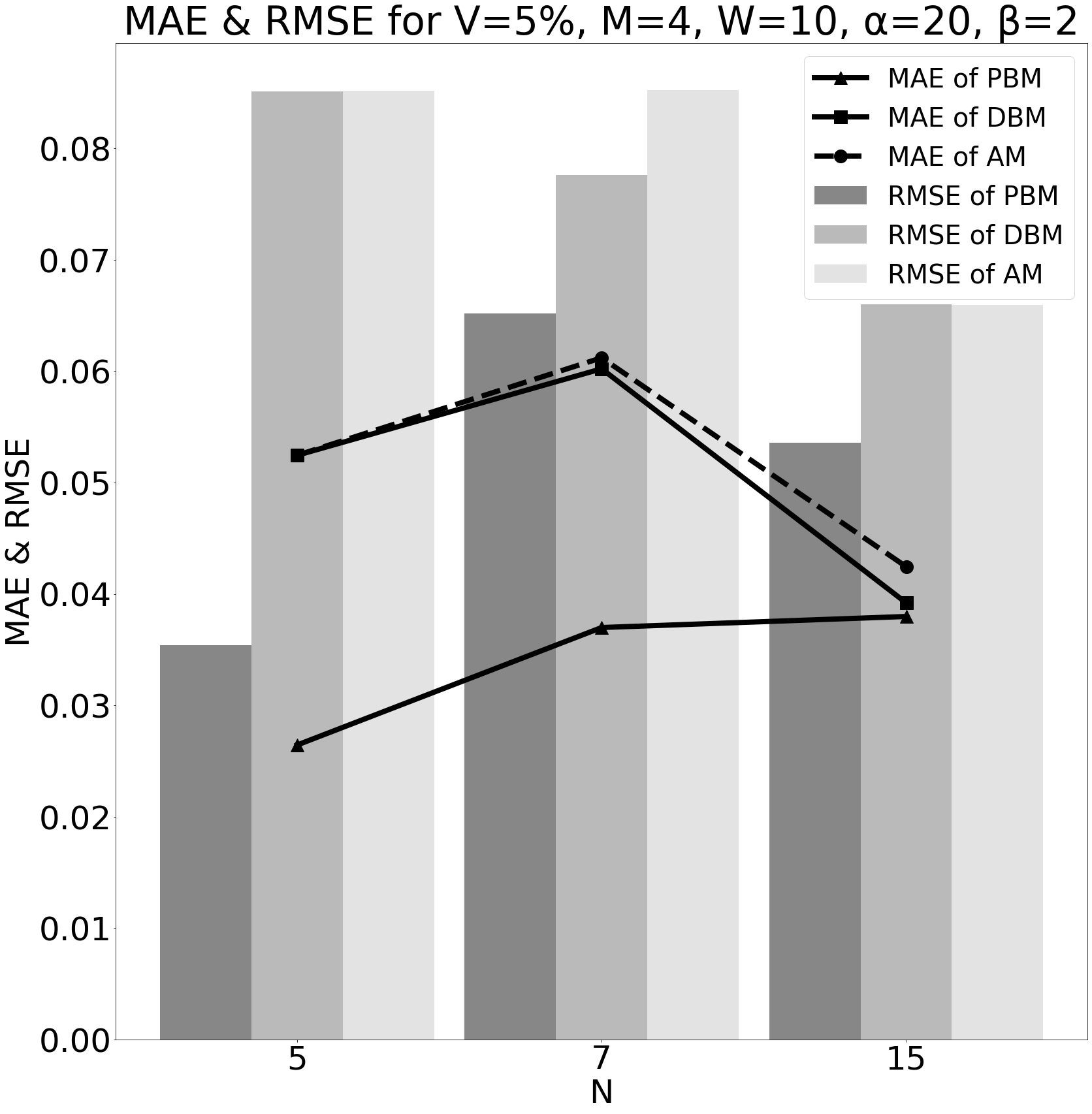}
	\caption{MAE and RMSE for $V = 5$\% and $M = 4$.}
	\label{fig4}
\end{figure}

\begin{figure}[!h]
	\centering
	\includegraphics[scale=0.10]{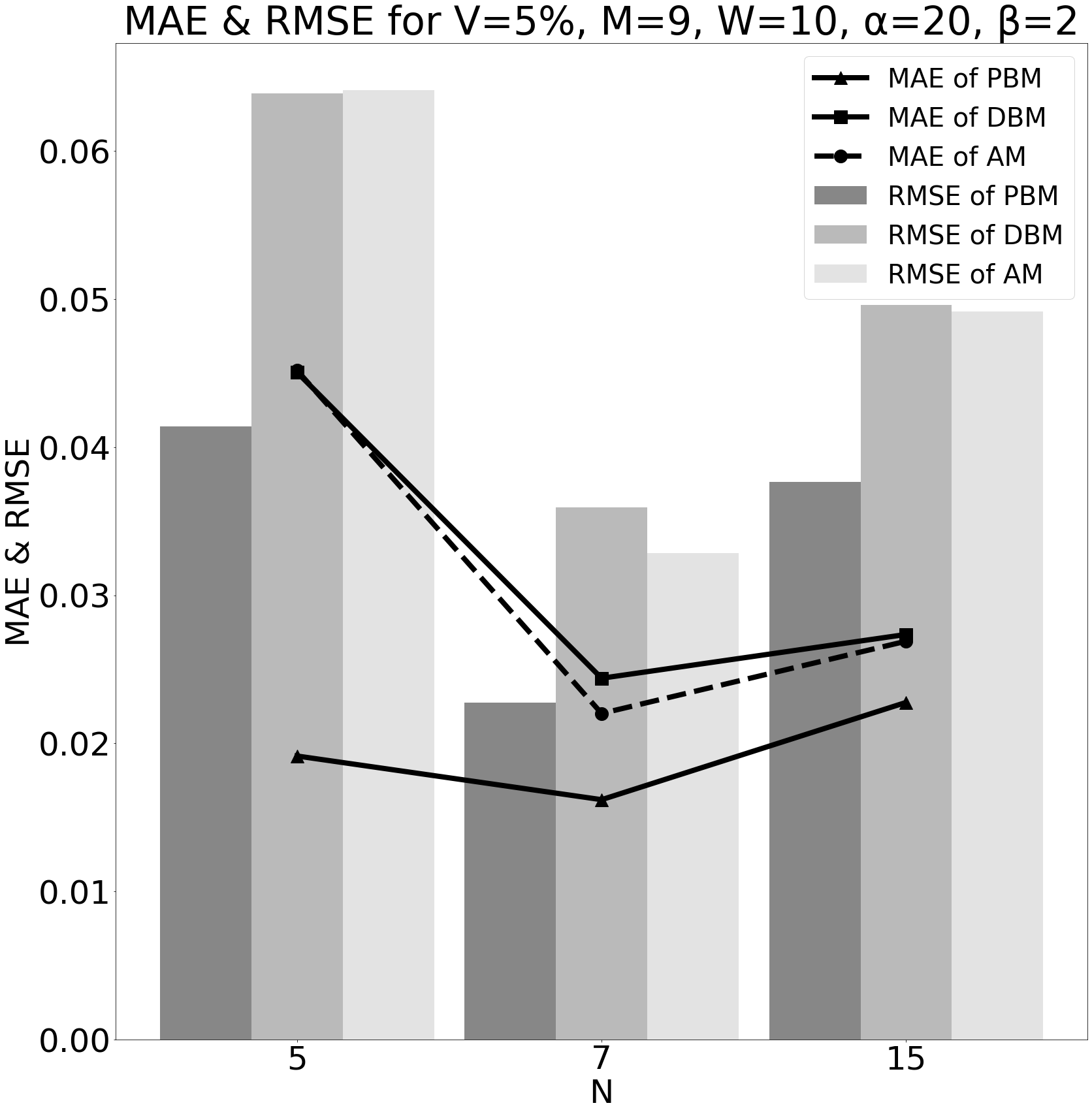}
	\caption{MAE and RMSE for $V = 5$\% and $M = 9$.}
	\label{fig5}
\end{figure}

In Figures \ref{fig4} \& \ref{fig5}, we present our results for $V = 5$\% and $M \in \left\lbrace 4, 9 \right\rbrace$. Now, we increase the number of missing values present into our datasets, i.e., devices' reports.
In this set of experiments, the PBM clearly outperforms the remaining models no matter the number of dimensions and devices. The PBM manages to achieve a very low error when called to provide the replacements.
An interesting observation is that the PBM is negatively affected by $M$ \& $N$ while the DBM and the AM are positively affected by the same parameters. In any case, the difference in the performance is high when we focus on a low number of dimensions and devices.

\begin{figure}[!h]
\centering
	\includegraphics[scale=0.10]{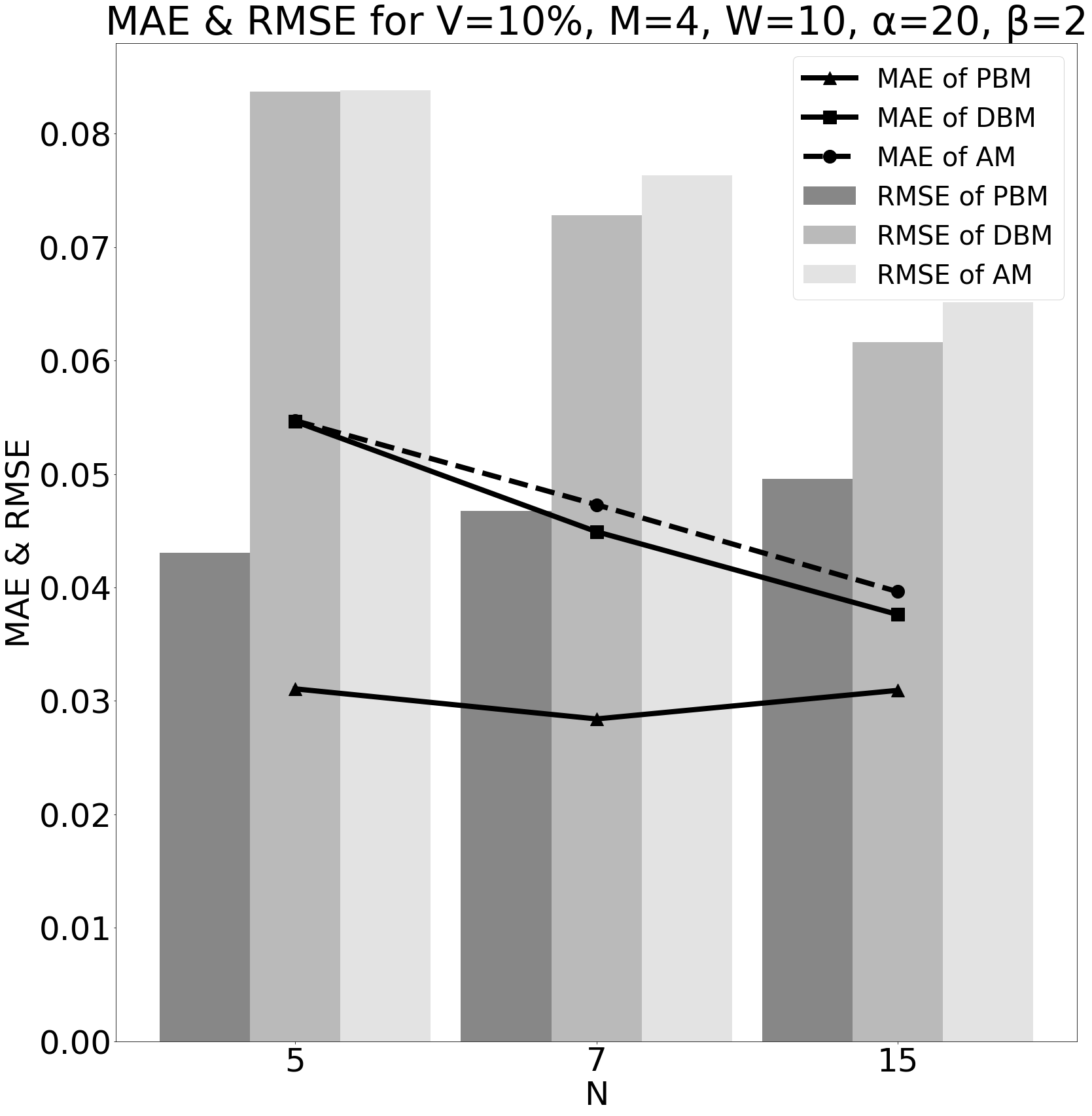}
	\caption{MAE and RMSE for $V = 10$\% and $M = 4$.}
	\label{fig6}
\end{figure}

\begin{figure}[!h]
\centering
	\includegraphics[scale=0.10]{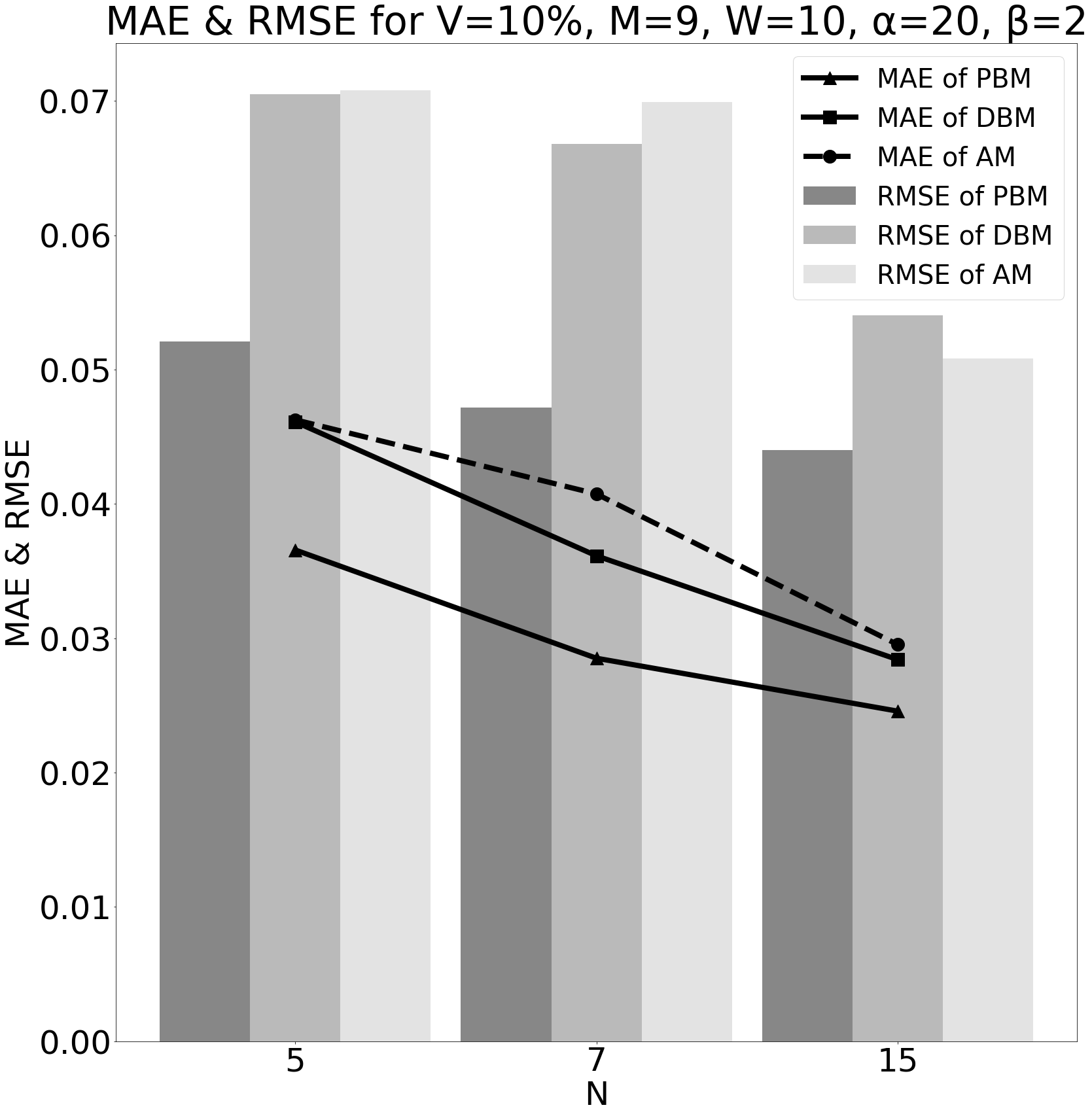}
	\caption{MAE and RMSE for $V = 10$\% and $M = 9$.}
	\label{fig7}
\end{figure}

In Figures \ref{fig6} \& \ref{fig7}, we increase the number of missing values assuming $V = 10$ and provide our experimental outcomes for 
$M \in \left\lbrace 4, 9 \right\rbrace$.
We observe a similar performance as in the previous experimental scenario. Current results conform the ability of the PBM to outperform in case of a high number of missing values. 
The error achieved by the PBM faces a slight increment when $M=4$ and is decreasing when $M=9$ (always considering that $N$ increases from 5 to 15 devices). The remaining models exhibit worse performance than the PBM in all the experimental scenarios. 

\begin{figure}[!h]
\centering
	\includegraphics[scale=0.088]{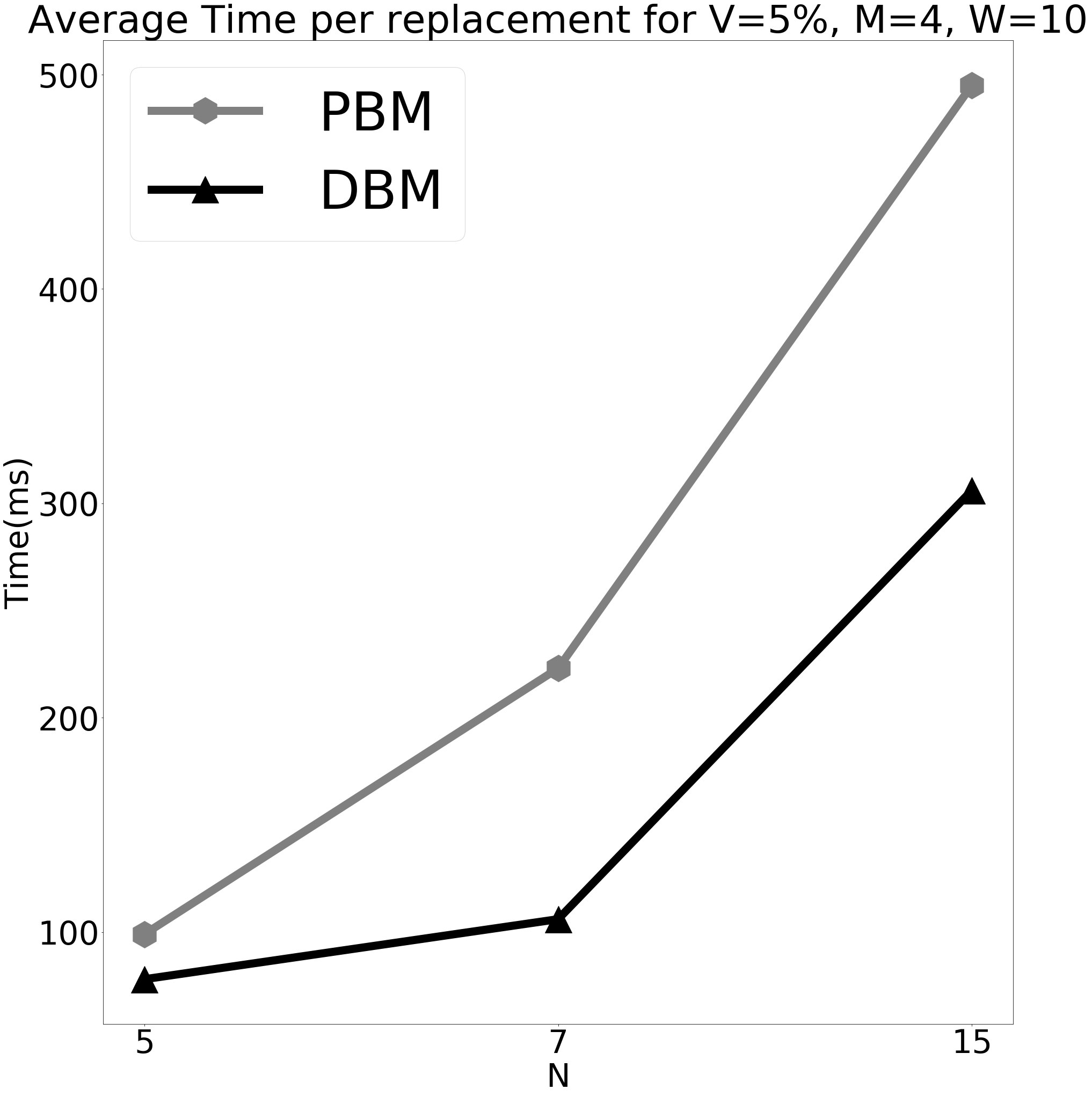}
	\caption{Time requirements for $V = 5$\% and $M = 4$.}
	\label{fig8}
\end{figure}

\begin{figure}[!h]
	\centering
	\includegraphics[scale=0.088]{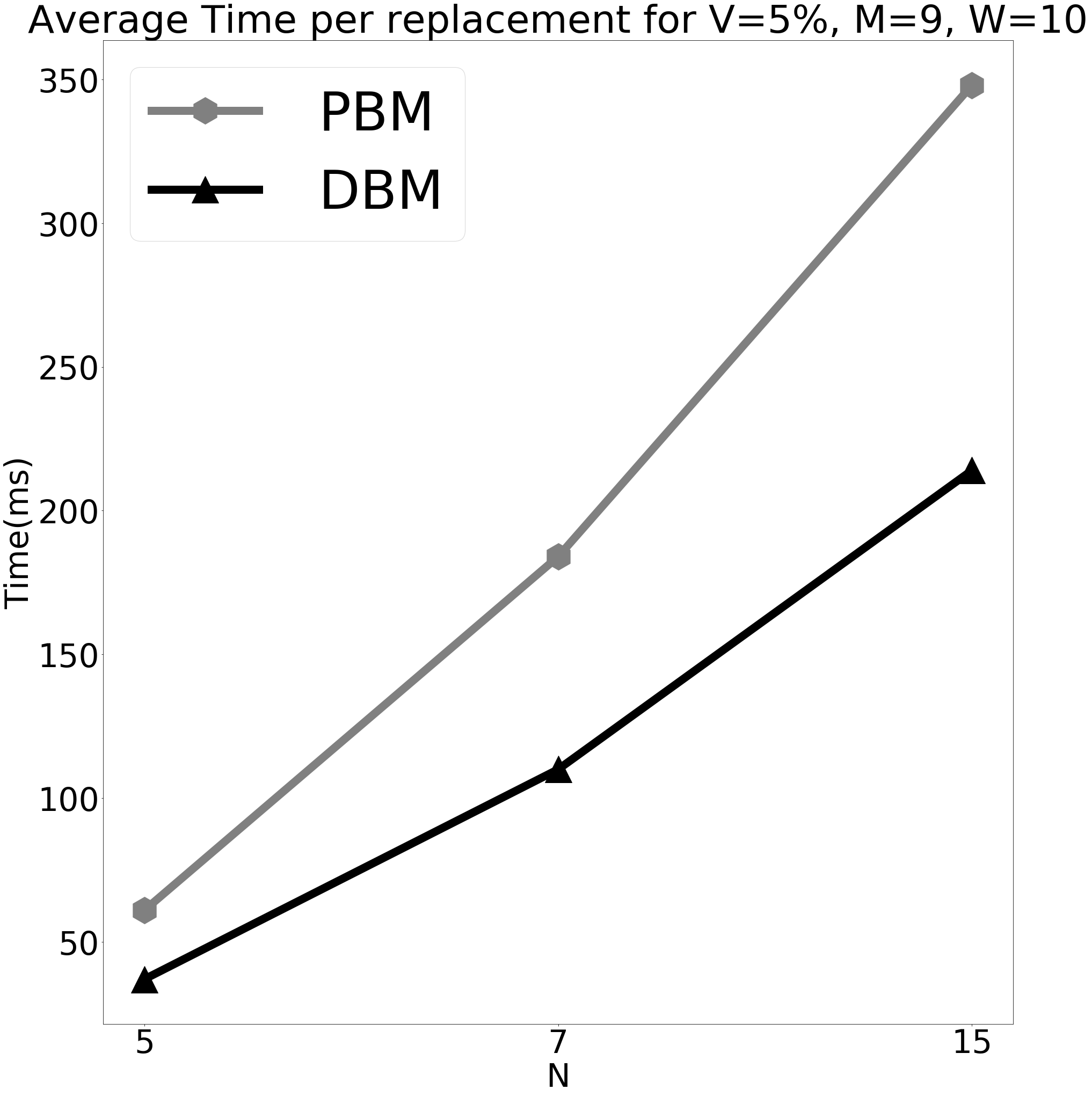}
	\caption{Time requirements for $V = 5$\% and $M = 9$.}
	\label{fig9}
\end{figure}

In Figures \ref{fig8} \& \ref{fig9}, we present our results related to the mean time required to calculate the final replacement for each  missing value. in this set of experiments, we get $V = 5\%$ and $M \in \left\lbrace 4, 9\right\rbrace$, respectively.
As we can observe, the number of the devices negative affect the mean time per missing value for both for PBM and DBM. This is natural, as we have to process more streams with a clear impact on the conclusion time. 
When we consider the PBM performance, we observe that it has better results for error metrics than the remaining models, however, it requires more time to conclude the replacements. In the worst case, the
PBM requires 500 ms to conclude a replacement
(compared to 300 ms of the DBM), i.e., it exhibits a throughput of two replacements per second. In any case, these outcomes reveal that the PBM can be adopted to support real time applications with a significant improvement in the performance (as our results for the delivered error exhibit) compared to the remaining models. It becomes obvious the trade off between the improved performance and the increased calculation time. We can tolerate an increment in the adopted calculations, thus, the required time upon the elimination of the error in the prediction of the appropriate replacements. 

We have to notice that the PBM requires a `warm up' period, up to $W$, to collect the necessary data and 
perform the envisioned calculations. This step is necessary to feed our distance/similarity and regression schemes. Additionally, a potential limitation concerns the 
scenario where edge nodes collect multiple missing values from multiple IoT devices at the same time instance. 
In this case, our similarity models cannot efficiently conclude the replacements as the final similarity may be wrongly concluded 
due to the limited dimensions participating in the above described calculations. In any case, we consider such a scenario as rare to be met in real applications. 

\section{Conclusion and Future Work}
\label{section6}
When focusing on applications over the IoT infrastructure, 
we have to adopt effective data imputation techniques 
that are capable of providing the final result in the minimum time.
The reason is that we have to efficiently support novel applications with significant impact to end users due to the closeness that IoT devices exhibit with them. This way, we can keep many computational tasks close to users 
to eliminate the latency faced in applications' responses.
Moreover, IoT devices are directly connected with the infrastructure 
at the edge of the network.
Edge nodes act as intermediaries between 
IoT devices and the Cloud becoming an additional point where the collected can be processed.
In this paper, we propose a data imputation technique to be adopted 
at edge nodes.
We consider the monitoring of data reported by IoT devices and 
an efficient mechanism for calculating replacements for missing values.
Our model builds on multiple schemes creating an ensemble approach.
We rely on the spatio-temporal aspect of the problem and propose replacements upon the view of each device (as defined by its historical observations - the `local' view) and the view of a group of devices exhibiting a high similarity with the device reporting a missing value.
Widely known metrics like the cosine similarity 
and the Mahalanobis distance are smoothly combined to support the replacement proposed by the group of peer devices. 
A regression based model is also adopted to 
deliver the replacement based on the `local' historical observations.
Both inputs, i.e., (the group and the local views) are `aggregated' based on a dynamically adapted scheme that defines the weight for each replacement.
Our technique manages to find the appropriate replacements for each missing value as exposed by the prediction error and our comparative assessment.
Our performance evaluation process reveals the ability of the proposed scheme to outperform other similar models upon different real traces and the limited time for concluding the envisioned calculations. 
Our future research plans involve 
the definition and adoption of a more 
complex methodology taking into consideration the 
uncertainty behind the involvement of specific peer devices in the
envisioned processing.

\end{document}